\documentclass{elsart}
\usepackage[xdvi]{graphicx}
\usepackage{amssymb}
\usepackage{pstricks,pst-node,pst-text}
\newcommand{\beq}{\begin{center}\begin{equation}}
\newcommand{\eeq}{\end{equation}\end{center}}

\newcounter{bla}

\begin{document}
\bibliographystyle{cpc}
\begin{frontmatter}

\title{Optimized Multiple Quantum MAS Lineshape Simulations in Solid State NMR }

\author[a]{William J Brouwer\thanksref{author}},
\author[a]{Michael C Davis},
\author[a]{Karl T Mueller}

\thanks[author]{Corresponding author}

\address[a]{ Department of Chemistry, Pennsylvania State University }

\begin{abstract}
The majority of nuclei available for study in solid state Nuclear Magnetic Resonance have half-integer spin $I > 1/2 $, with corresponding electric quadrupole moment. As such, they may couple with a surrounding electric field gradient. This effect introduces anisotropic line broadening to spectra, arising from distinct chemical species within polycrystalline solids. In Multiple Quantum Magic Angle Spinning (MQMAS) experiments, a second frequency dimension is created, devoid of quadrupolar anisotropy. As a result, the center of gravity of peaks in the high resolution dimension is a function of isotropic second order quadrupole and chemical shift alone. However, for complex materials, these parameters take on a stochastic nature due in turn to structural and chemical disorder. Lineshapes may still overlap in the isotropic dimension, complicating the task of assignment and interpretation. A distributed computational approach is presented here which permits simulation of the two-dimensional MQMAS spectrum, generated by random  variates from model distributions of isotropic chemical and quadrupole shifts. Owing to the non-convex nature of the residual sum of squares (RSS) function between experimental and simulated spectra, simulated annealing is used to optimize the simulation parameters. In this manner, local chemical environments for disordered materials may be characterized, and via a re-sampling approach, error estimates for parameters produced. 



\end{abstract}

\end{frontmatter}

\section{Introduction}

Since the discovery of Nuclear Magnetic Resonance (NMR), there has been great interest in the study of quadrupolar nuclei. These nuclei have an electric quadrupole moment $Q$, which couples with a non-zero electric field gradient. As a result, anisotropic frequency dependence is introduced, promoting overlap between lineshapes  arising from distinct chemical sites in powdered solids and degrading resolution. This issue has  been addressed over the course of time by a number of experimental approaches. Early in the development of solid state NMR, Magic Angle Spinning (MAS)~\cite{and59} was proposed, which reduces or eliminates second rank interaction terms and therefore broadening associated with the first order quadrupole interaction. This interaction depends explicitly on the angle $\theta$ between sample rotor axis and the static, applied field of NMR. Attention here is restricted to first and second order quadrupole effects, each of which is a function of the second order Legendre polynomial $P_2(\theta)$. The second order quadrupole perturbation is also a function of the fourth order Legendre polynomial $P_4(\theta)$.  Additionally, an appreciable second order isotropic shift inversely proportional to the Larmor frequency $\omega_0$ occurs; the center of gravity of a quadrupole lineshape is subsequently changed from the chemically shifted value. The characteristic features of quadrupole spectra provide valuable local bonding information and hence extensive work has been devoted to both resolving individual chemical sites, as well as lineshape simulation. However, if the magnitude of the quadrupole interaction is significant, spinning sideband manifolds arising from satellite frequency transitions may still obscure spectra in one dimension~\cite{mar79}. Double Rotation (DOR)~\cite{sam88} and   Dynamic Angle Spinning (DAS)~\cite{ktm90,ktm91} are successful in eliminating the effects of both second and fourth rank tensor terms, and thus also second order quadrupole broadening. More recently, Multiple Quantum Magic Angle Spinning (MQMAS)~\cite{med95,fry95} and Satellite Transition Magic Angle Spinning (STMAS)~\cite{gan00,gan03,ash02} have become popular owing to mechanical simplicity. These procedures involve collecting data as a function of two independent time intervals in the pulse sequence~\cite{ern66} under Magic Angle Spinning conditions. Within the MQMAS experiment, directly observable single quantum coherence frequency transitions  are correlated with multiple quantum transitions~\cite{pia82} and, in the case of STMAS, satellite transitions,  which evolve between pulses and are selected via an appropriate phase cycle. From the center of gravity of peaks along the high resolution axis, isotropic shifts are deduced which are a function of both isotropic chemical ($\delta^{iso}_{cs}$) and second order quadrupole ($\delta^{iso}_{2Q}$) shifts. In turn, the isotropic second order quadrupole shift is a function of both the quadrupole coupling constant $C_q$ and asymmetry parameter $\eta_q$. The importance of these quantities lies in the fact that they are functions of the electric field gradient tensor ${\cal V}$, and thus the details of the local bonding environment:\beq \eta_q = \frac{{\cal V}_{yy}-{\cal V}_{xx}}{{\cal V}_{zz}}; \mbox{   }C_q = \frac{e {\cal V}_{zz} Q}{\hbar}. \eeq 

In order to unequivocally determine both $C_q$ and $\eta_q$, simulation of experimental spectra is necessary~\cite{mas02}.  In the case of disordered chemical environments~\cite{hoa02,bur992,cha98,hwa97}, calculations of powdered lineshapes for MQMAS becomes a formidable task. This is due to the fact that parameters relevant to simulation take on a distributed nature~\cite{bod98,czj81}.  The focus of this paper is devoted to the optimized simulation of multiple quantum magic angle spinning spectra, in the presence of low to significant disorder. This is accomplished using quasi-random numbers sampled from model distributions of isotropic chemical shift and quadrupole coupling constant. Simulated annealing is used to optimize the non-convex RSS function, and in distinction to existing simulation methods, model parameter error estimates are calculated, using the non-linear jackknife~\cite{fox80}. The overall process has been implemented in the C programming language with some tasks performed using the OCTAVE scripting language, and is highly amenable to distributed computing~\cite{kri03}.

\section{Theoretical background}
\subsection{Lineshape Simulation}
Since the introduction of MQMAS experiments, there have been significant improvements in excitation efficiency and coherence transfer, for example, using Double Frequency Sweep (DFS)~\cite{ken99,ken91} and Fast Amplitude Modulation~\cite{mad99,mad02}. There have also been improvements made in sensitivity based around the inclusion of signal intensity from additional coherence transfer pathways~\cite{gan04,mal05}. The Z-filter~\cite{amo96} method ensures that amplitudes for echo and anti-echo pathways are co-added with equal intensity under States~\cite{sta82} acquisition, providing after phase correction a purely absorptive 2-D spectra. Given these improvements, particularly the latter, it is reasonable to assume that lineshapes for individual crystallite orientations may be described via traditional linear response theory~\cite{kub54,how03,abra}. Further, allowing for the possibility of contributions from both homogenous and inhomogeneous broadening processes, a complete model includes a linear combination of Lorentzian and Gaussian absorption lineshapes with broadening factors $\lambda_2,\lambda_1$:
\[ F(f1,f2)=(1-\epsilon)\frac{\lambda_1}{\lambda_1^2+(f1-f1_m)^2}\frac{\lambda_2}{\lambda_2^2+(f2-f2_m)^2} \]
\beq +\epsilon\frac{1}{2\pi\lambda_1\lambda_2}e^{\left(\frac{-(f1-f1_m)^2}{2\lambda_1^2}+\frac{-(f2-f2_m)^2}{2\lambda_2^2}\right)}  \eeq
where $\epsilon < 1$ is a free parameter, describing the relative fraction of different lineshape functions. It is assumed for the remainder of this work that attention is restricted to symmetric transitions (eg., 3QMAS experiments) and thus devoid of first order quadrupole effects, or that first order effects are absent from satellite transitions, the latter ensured by using an accurately set magic angle. Finally, it is assumed that experiments are conducted using a rotor-synchronized F1 dimension to eliminate spinning sidebands in this dimension~\cite{mas296}. Under these assumptions, the indirect $2\pi f1_m = \omega_{r,c}^{(2)}$ and directly  detected frequencies $2\pi f2_m = \omega_{-1}^{(2)}$  have the general form\footnote{Frequency transitions are labeled by $r$ and $c$. Owing to the dipole selection rule, directly detected frequency transitions are always such that $r-c=\pm 1$ eg., the central transition ($-1/2 \leftrightarrow 1/2$). Multiple quantum transitions are such that $r-c \neq \pm 1$ eg., the triple quantum transition  ($-3/2 \leftrightarrow 3/2$). The particular multiple quantum transition(s) correlated with the central transition in the course of an experiment are determined by the phase cycle.}
\[\omega_{r,c}^{(2)}=(r-c)\omega_0\delta^{iso}_{cs}-\frac{r-c}{\omega_0}\Omega_Q^2\left\{A^{(0)}(I,r,c)\left(\frac{{\eta_q^2+3}}{10}\right) \right. \] \beq + \left. A^{(4)}(I,r,c)f(\eta_q,\alpha,\beta)\right\}, \eeq

where:

\[ A^{(0)}(I,r,c)=I(I+1)-3(r^2+rc+c^2) \]
\beq A^{(4)}(I,r,c)=18I(I+1)-34(r^2+rc+c^2)-5 \eeq

are spin ($I$) and quantum transition ($r,c$) dependent constants. The isotropic, second order quadrupole shift $\delta^{iso}_{2Q}$ is given by the second set of terms in equation 3 divided by the Larmor frequency $\omega_0$, and contains the quadrupole coupling constant implicitly:
\beq {\Omega_Q}=\left[\frac{e{\cal V}_{zz}Q}{2I(2I-1)\hbar}\right] = \frac{C_q}{2I(2I-1)}\eeq

The second order, quadrupolar line broadening is described by function \\ $f(\eta_q,\alpha,\beta)$~\cite{man97}. This term is a function of the asymmetry parameter $\eta_q$ and powder angles $\alpha,\beta$, the latter describing the orientation between the Principal Axis System (PAS) of the electric gield gradient tensor and the rotor fixed frame:
\[f(\eta_q,\alpha,\beta) = \frac{1}{15120}[(-54-3\eta_q^2+60\eta_q\cos 2\alpha - 35\eta_q^2\cos 4 \alpha) \]
\[+(540 + 30\eta_q^2-480\eta_q\cos2 \alpha + 70\eta_q^2\cos 4\alpha)\cos^2\beta \]
\beq (-630-35\eta_q^2+420\eta_q\cos 2\alpha - 35\eta_q^2 \cos 4\alpha)\cos^4\beta ]\eeq

This quantity is a direct consequence of the transformation between principle axis frame of the crystallite and rotor fixed frame, in terms of Wigner rotation matrices. Assuming experiments are conducted in the fast MAS limit, where attention may be restricted to the centerband, a third angle $\gamma$ describing the rotor orientation with respect to the static field is unnecessary, since the static field represents a symmetry axis for spins. Throughout the course of an MQMAS or STMAS experiment, or via subsequent data processing, the indirect dimension frequency $f1_m$ becomes $f1_m' = f1_m - k \times f2_m$, the shearing factor $k$ often chosen to eliminate the anisotropic frequency component and thus create a fully isotropic frequency dimension. The resultant frequency $f1_m'$  as well as the accompanying bandwidth may be rescaled by a factor $1/(1+k)$, according to one convention. For ease of comparing spectra arising from different multiple quantum experiments, this work follows the unscaled representation~\cite{man98}. Note that regardless of the convention followed in presentation and analysis of spectra, the isotropic chemical shifts ultimately deduced are identical.  Equation 2 is germane to a single crystallite orientation, with a particular isotropic chemical shift, asymmetry parameter and quadrupole coupling constant. A more general lineshape intensity function for a powdered solid must be weighted by crystallite angle distribution $G(\alpha,\beta)$. In addition, in the presence of disorder, the experimental lineshape is averaged due to distributed values of $\delta_{cs}^{iso},C_q,\eta_q$, described by probability density  $P(\delta_{cs}^{iso},C_q,\eta_q)$:
\[ I(f1,f2) = \] 
\beq\sum^M_i{\cal A}_i\int_{\delta_{cs}^{iso},C_q,\eta_q}\int_{\alpha,\beta} P_i(\delta_{cs}^{iso},C_q,\eta_q) G(\alpha,\beta) F_i(f1,f2) d\alpha d\beta d[\delta_{cs}^{iso},C_q,\eta_q]_i \eeq

where $M$ is the total number of chemical sites and ${\cal A}_i$ the individual site amplitude. There are two basic aspects to a numerical evaluation of this five dimensional integral, including powder averaging over the crystallite orientations. In addition, contributions to the overall spectrum from random variates $C_q, \delta^{iso}_{cs},$ and $\eta_q$ are weighted by a multi-variate probability distribution function, distinct for each site. The former aspect, powder averaging in magnetic resonance, is an example of a problem in broader quantum mechanics, evaluating integrals over the unit sphere~\cite{fre80,kuo05}. There exist several reviews in the literature with regard to powder averaging in magnetic resonance~\cite{pon99,hod00}. It is assumed here that the equally probable crystallite orientations within a powder have been equally irradiated, and the integral over angles is replaced by a sum:
\beq \bar F(f1,f2) = \frac{\sum_k w_kF_k(\alpha,\beta)}{\sum_k w_k} \eeq

 with various choices for weights $w_k$ and angles $\alpha,\beta$. Under this assumption, the contribution of a particular crystallite orientation to the overall intensity is proportional to $d\alpha d\beta \sin \beta$ ie., $G(\alpha,\beta) = \sin \beta$. The particular powder integration scheme used within this work corresponds to the Zaremba-Conroy-Wolfsberg (Z-C-W) method~\cite{zar66,con67,che73}, where angles and weights are chosen according to:\beq \begin{array}{ll}

\alpha_k= &\frac{2\pi(kM_a\mbox{ mod }N_a)}{N_a} \\
\beta_k= & \arccos\left(1-\frac{2k+1}{N_a}\right) \\
w_k= & 1 \\
\end{array} \eeq

with $N_a$ and $M_a$ chosen to satisfy $M_a=F(n)$ and $N_a=F(n+2)$, where $F(n)$ is the $n$th Fibonacci number, and index $k=0,1,...,N_a-1$ . This particular approach is considered preferable under fast MAS conditions~\cite{hod00} and demonstrates very good convergence versus order $n$.  

The second major aspect to evaluating equation 7 involves averaging over isotropic chemical shift and quadrupole parameters,  accomplished via Monte Carlo simulation. In general, statistical distributions may be symmetric or asymmetric. The nature of the model distribution used in the simulation is directly related to the underlying chemical and/or structural disorder.  Traditional random number generators which create variates according to probability distributions  are usually one of two types. They may be of the acceptance/rejection type, or rely on transformations of the uniform distribution, eg., the Box-Muller method for normal-distributed variables~\cite{nie92}. The latter was used here for ease of adaptation to a parallel programming environment. By creating Gaussian distributed variates, the integral of eq. 7 over the probability distribution may be converted to a summation, and the powder-averaged kernel $\bar F(f1,f2)$  simply evaluated as a function of the variates. By the law of large numbers, Monte Carlo approximations converge to the true value in the limit as the samples $N$ approach infinity. In reality, convergence is slow, and the error in using pseudo random numbers is $O(N^{-1/2})$. This situation is improved via using quasi-random numbers such as the Sobol sequence, which have an error $O((\log N)^m N^{-1})$ for $m$ dimensions~\cite{caf94}. For the purposes of this work, attention is restricted to  the bi-variate ($m$=2) Gaussian distributions in $\delta^{iso}_{cs}$ and $C_q$  (whose random variates are represented by $x$ and $y$ respectively):

\beq P(x,y) = e^ {- {{{{\left(y-\mu_y \right)^2}\over{\sigma_y^2}}-{{2\,\rho\,\left(x-\mu_x \right)\,\left(y-\mu_y\right)}\over{\sigma_x\, \sigma_y}}+{{\left(x-\mu_x\right)^2}\over{\sigma_x^2}}}\over{2\,\left(1-\rho^2\right)}} } \eeq

 For each chemical site, this distribution is parameterized by site-specific values for $\mu_x,\sigma_x,\mu_y,\sigma_y,\rho$,  where $\rho$ is the correlation coefficient between chemical shift and quadrupole coupling constant only. At this stage, single values for $\eta_q$ were deemed sufficient to model lineshapes. This was due to an observed insensitivity of lineshape simulation to a range of values for $\eta_q$. To summarize thus far then, each chemical site $i$ is modeled using ten free parameters ${\bf a}^i$:

\beq  {\bf a}^i =\{\lambda_1^i,\lambda_2^i,\epsilon^i,\mu_x^i,\sigma_x^i,\mu_y^i,\sigma_y^i,\rho^i,\eta_q^i,{\cal A}^i\};i=1,..,M\eeq

The integral of (7) is replaced by a double summation, in performing powder angle and parameter averaging tasks:

\beq \frac{1}{N\cdot (N_a-1)} \sum^M_i {\cal A}_i \sum_{x,y}^N \sum_{\alpha_k,\beta_k}^{N_a-1} F_i(f1,f2) \eeq
where $N$ is the total number of variates $x,y$ for each chemical site $i$. These variates are sampled from a bi-variate Gaussian distribution, using the Box-Muller transformation of (Sobol) quasi-random numbers on [0,1). The powder angles $\alpha,\beta$  are chosen according to the Z-C-W scheme, as is the number of summands $(N_a-1)$.

\subsection{Optimization}
Using the theory outlined thus far, an experimental spectrum may be simulated and attempts made to optimize the simulation parameters. In reality, modeling the underlying parameter distributions implies that at least two chemical sites are used in the optimization. Figure 1 is a plot of the RSS function obtained by varying only chemical shifts in an optimization for a two site MQMAS spectrum.

\begin{figure}[h!]
\begin{center}
\scalebox{0.5}{\rotatebox{0}{\includegraphics{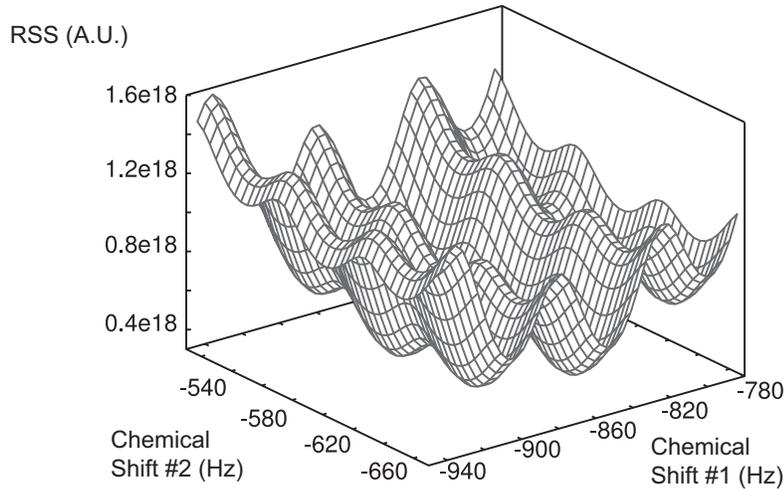}}}
\caption{RSS function, sum of squared difference between simulated and experimental MQMAS spectrum, as a function of the two isotropic chemical shifts.}\end{center}
\end{figure}

The surface is highly non-convex; the global minima is toward the center of the plot, within a larger area containing local minima. Simulated annealing~\cite{kir83} is a stochastic method for global optimization highly suited to non-convex RSS functions. The method is analogous to the metallurgical process of annealing. The application to the current problem ensures that the iterative procedure avoids being trapped within local minima; the overall algorithm applied here is as follows:

\begin{enumerate}
\item RSS function or generalized energy generation, the trace of the Grammian:
 \[E_0 = \mbox{ Trace }\{(A-B)\times(A-B)^T\}\]
 where $A-B$ is a matrix of residuals, the difference between simulated $A$ and experimental absorption spectra $B$. If this is the initial step, a generalized temperature is defined $T \approx E_0$
\item Each unconstrained parameter ${\bf a}^i$ is changed by a random amount $\pm p\Delta {\bf a}^i$,  $p$ sampled from the uniform distribution $[0,1)$. The corresponding energy $E_f$ is calculated as before.   
\item If $E_f<E_0$, the change is accepted, else,
\item Parameter changes are accepted or rejected in the traditional Metropolis~\cite{met53} scheme, using the probabilistic factor: $e^{-(E_f-E_0)/T}$ 
\item The process is repeated and the temperature lowered according to some schedule, until such time as convergence is reached. 
\end{enumerate}   

Implicit to the algorithm is the need to choose an appropriate maximum step size $\Delta {\bf a}^i$ and annealing schedule. To ensure adequate search of the parameter space, $\Delta {\bf a}^i$ was fixed between one and two percent of the starting parameter values. The annealing schedule is more subjective and best determined via experiment. A common method involves reducing the temperature at every step by some amount $\delta$:

\beq T_f = (1-\delta)T_0 \eeq

which requires the tuning of $\delta$. Significant gains are made during the early stages of the algorithm, during which there is a non-zero probability for energy to increase. In order to exploit this feature, $\delta$ was set to approximately 0.5 and the schedule of equation 13 was re-set every $\kappa$ steps to the current best value of energy, a process of rapid annealing and re-annealing.

\subsection{Error Estimation}
In order to give confidence intervals for the free parameters listed in eqn. 11 optimized in the simulation, strictly speaking the measurement or MQMAS experiment in conjunction with simulations ought to be repeated and statistics created from fitted data. However, owing to the considerable time multiple experiments and simulations requires, a more suitable approach to error analysis is found in statistical re-sampling~\cite{leo97}, such as jackknifing or bootstrapping~\cite{shao}. In the original jackknife approach, $\bar\phi_{-j}$ is defined as the least squared estimate of parameter $\phi$ when the $j$th data point of $n$ total is removed from the set.  Pseudo values are created,
\beq P_j=n\bar\phi-(n-1)\bar\phi_{-j} \eeq
with average $\bar P$ and variance matrix $V_P$: 
\beq \bar P=\bar\phi_J=n^{-1}\sum_{j=1}^{n} P_j \eeq
\beq nV_P = \frac{1}{n-1}\sum_{j=1}^{n}(P_j-\bar P)(P_j-\bar P)^T \eeq

In the present application, this method implies $n+1$ non-linear optimizations which is still far too time consuming. Fox et al~\cite{fox80} propose a solution in the form of an approximate jackknife, which requires instead a single non-linear optimization, via a Taylor expansion of the least squares estimate equation for $\bar \phi_j$, assuming it is a stationary point for the sum of the residuals. In this method, an estimate of the variance matrix $V_J$ is given by:
\beq V_J=(Z^TZ)^{-1}\sum_{j=1}^n z_jz_j^Tr_j^2(Z^TZ)^{-1} \eeq
where:
\beq z_j = \nabla f(x_j,\phi) = \left\{\frac{\partial}{\partial \phi_1} f(x_j,\phi)...\frac{\partial}{\partial \phi_l} f(x_j,\phi)\right\}_{\phi=\bar \phi}^T \eeq
\beq Z^T = (z_1,..., z_n) \eeq

and $r_j$ is the vector of residuals. The model as presented here consists of ten free parameters per chemical site (ie., $l$=10), so in the case of $M$ chemical sites, this corresponds to the creation of a $10M\times10M$ variance matrix. This matrix is evaluated at best-fit parameters $\bar {\bf a}^i$, using the partial derivatives of equation 7, listed in appendix A and evaluated as before via summation.

\section{Implementation}
The aforementioned theory was implemented in C, using a number of functions from the GNU Scientific Library (GSL), as well as the math and standard libraries. A single application was written which performs calculations of frequency equation 3, for each dimension. Further, a multiple $k$ of the direct dimension frequency $f2_m$ is subtracted from the indirect dimension frequency $f1_m$, according to the function of the shearing transformation. As stated previously, the shear factor $k$ is often chosen to produce a fully isotropic frequency dimension and for the examples given here (spin 5/2 and 3QMAS experimental conditions) corresponded to a numerical value of 19/12. For each frequency dimension, Sobol sequences are generated and used to create bi-variate distributions of isotropic chemical shift and quadrupole coupling constant according to the Box-Muller algorithm. Powder angles are generated according to the Z-C-W algorithm. Finally, summation over powder angles, variates and chemical sites are performed using equation 12. A single OpenMP pragma was used to parallelize inner frequency loops,

{\tt \#pragma omp parallel for private(h,i) }

using the private declaration on loop indices to prevent a race condition occurring between separate threads. The OpenMP application programming interface is  essentially a set of libraries and associated compiler directives which permits shared memory processing (SMP) on machines with the appropriate hardware. The C source was compiled using the GNU C compiler, linking the appropriate libraries:

{\tt gcc -O4 -o mqmas\_opt mqmas\_opt.c -lm -lgslcblas -lgsl -fopenmp}

In order to perform optimization of the simulation parameters, the simulated annealing algorithm was implemented in an OCTAVE script, {\tt mqmasOpt.m}. This allowed for tuning of heuristic parameters, particularly the annealing schedule and size of random fluctuations taken by individual parameters per iteration (set to between 2-3\% of initial parameter magnitudes). In addition, parameter values corresponding to the lowest energy obtained are stored every iteration and used for occasional resets.

\begin{figure}[h!]
\begin{center}
\scalebox{0.6}{\rotatebox{0}{\includegraphics{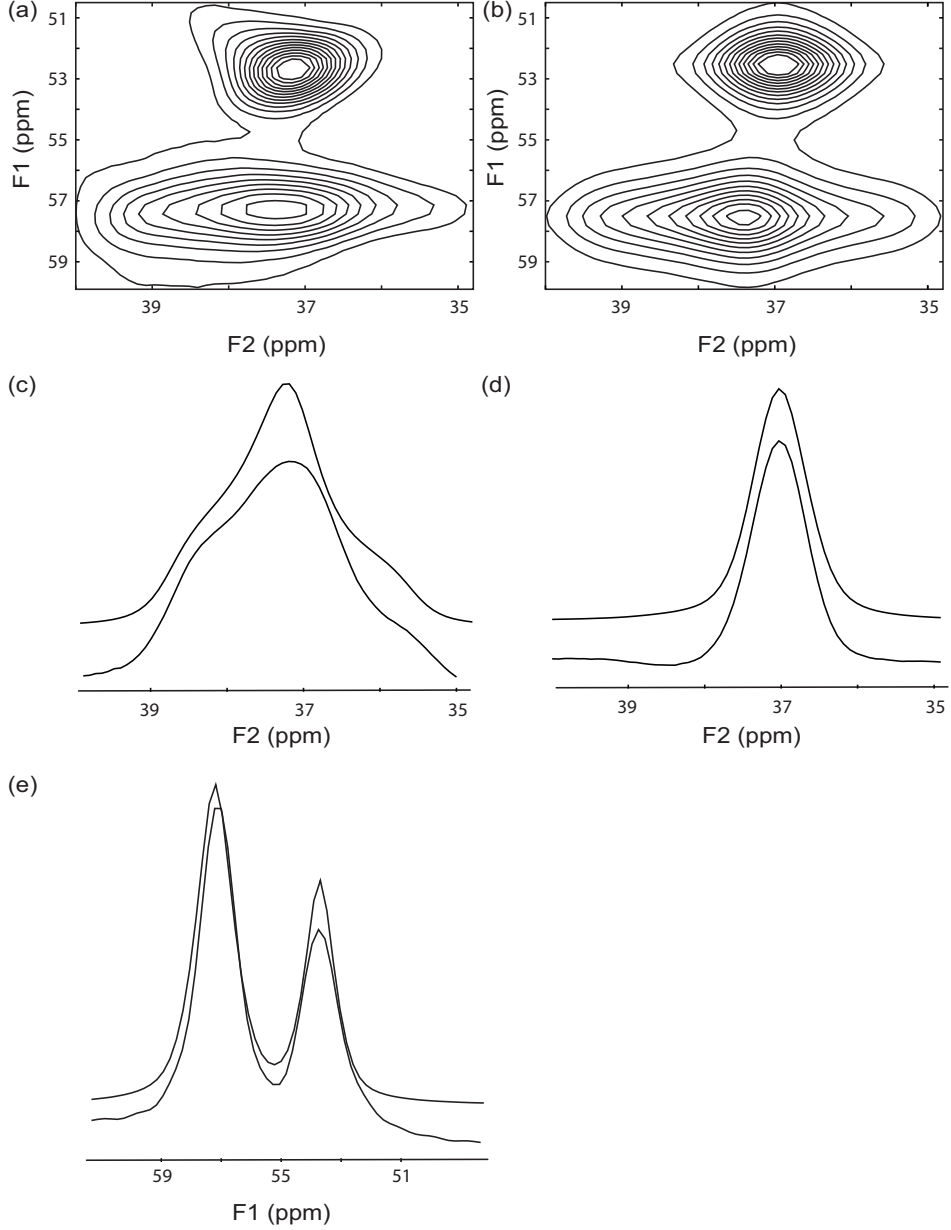}}}
\caption{ Spectra are referenced to AlCl$_3$ and scaled according to the second convention of Amoureux et al~\cite{amo98}, where indirect dimension bandwidth reflects spinning speed (a) $^{27}$Al 3QMAS VPI-5 spectrum obtained at 11.7T, using a Chemagnetics spectrometer and z-filter/States sequence~\cite{amo96}; spinning speed 10kHz, bandwidth 10kHz in each dimension, 64$\times$1024 total points in F1 and F2 respectively. (b) Simulation of experimental 3QMAS spectrum (c)Trace along frequency $f1$ = 58 ppm showing experimental (lower) and simulated (upper) spectrum (d) Trace along frequency $f1$ = 53 ppm showing experimental (lower) and simulated (upper) spectrum (e) Integrated intensity along isotropic dimension showing experimental (lower) and simulated (upper) spectrum. }

\end{center}\end{figure}

It is anticipated that the number of crystallite orientations required for adequate convergence in a particular simulation will increase with linewidth, which in turn is proportional to the quadrupole coupling constant. Fitting to a crystalline model compound provides a good means of determining the minimum number of crystallite orientations required for a comparable linewidth. Convergence or lack thereof is more easily observed in a crystalline system as compared to a more disordered material, which is devoid of the characteristic features. In order to test convergence of the powder averaging step for a material of interest, a $^{27}$Al 3QMAS spectrum of large-pore aluminophosphate VPI-5 was acquired using a 11.7T spectrometer, figure 2. To simulate the full 3QMAS spectrum without visible irregularities, 1597 angle pairs ($F_{17}$) were minimal for quadrupole coupling constants in the range less than 4 MHz, as exhibited by the model compound VPI-5. The Second Order Quadrupole Effect (SOQE) parameters\footnote{SOQE = $C_q\sqrt{1+\frac{\eta_q^2}{3}}$}  as determined from the simulation for the tetrahedral region of VPI-5 were 2.8 and 1.3 MHz, which compare favorably with literature values~\cite{roc96}. 

Using the same number of crystallite angles, optimized simulations were performed for the tetrahedral region within a hydrated albite sample, using 200 quasi-random samples for each of two chemical sites, drawn from two bi-variate distributions. Results are displayed in figure 3 and table 1.
\begin{figure}[h!]
\begin{center}
\scalebox{0.6}{\rotatebox{0}{\includegraphics{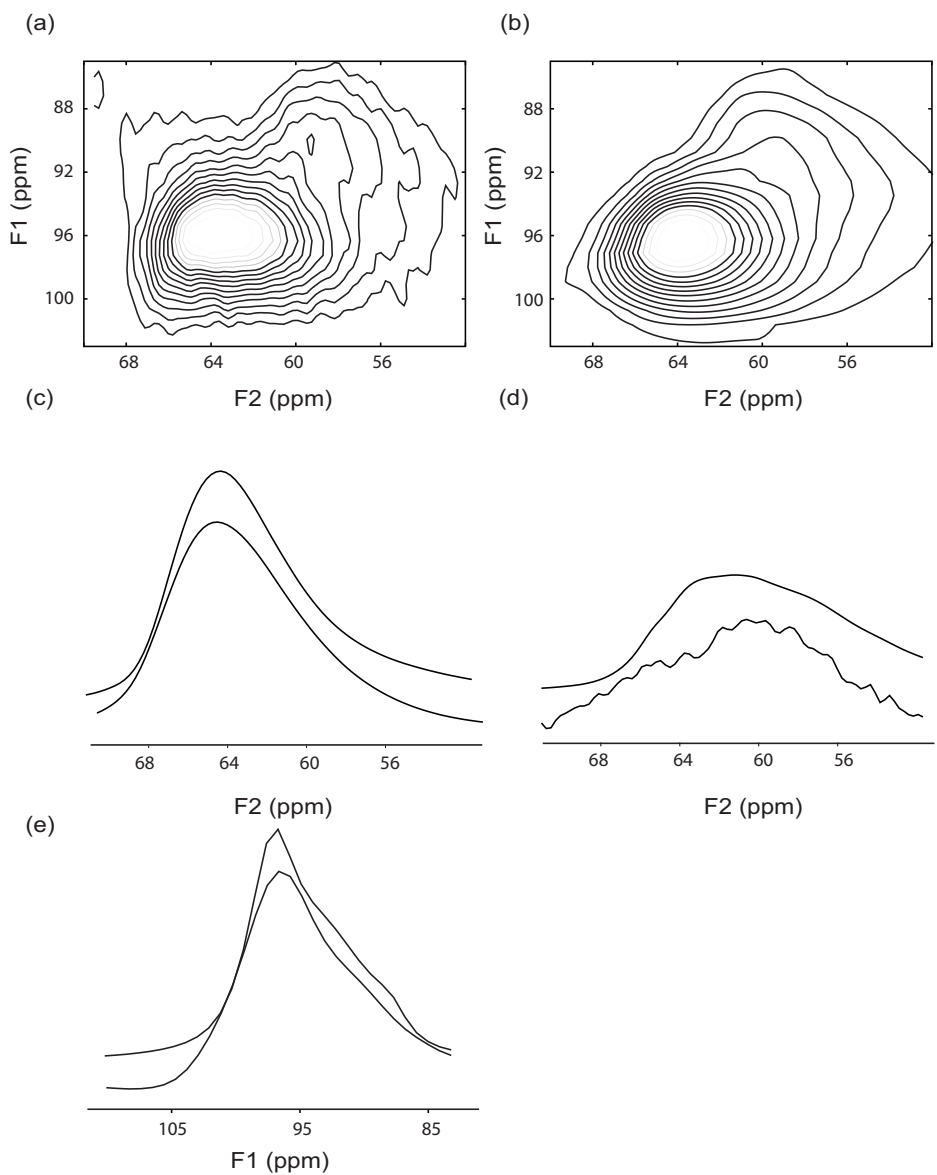}}}
\caption{Spectra are referenced and scaled as previously; (a) $^{27}$Al 3QMAS hydrated albite spectrum obtained at 11.7T, using a Chemagnetics spectrometer and z-filter/States sequence; spinning speed 10kHz, bandwidth 10kHz in each dimension, 64$\times$1024 total points in F1 and F2 respectively. (b) Simulation of experimental 3QMAS spectrum (c)Trace along frequency $f1$ = 96 ppm showing experimental (lower) and simulated (upper) spectrum (d) Trace along frequency $f1$ = 91 ppm showing experimental (lower) and simulated (upper) spectrum (e) Integrated intensity along isotropic dimension showing experimental (lower) and simulated (upper) spectrum.}\end{center}
\end{figure}

The Gaussian/Lorentzian ratio, correlation coefficient and broadening constants in both dimensions were constrained to 0.5, 0, and 100 Hz respectively and 1000 simulated annealing iterations were performed. The experimental spectrum displays regions of both order (narrow, horizontal peaks) and disorder (broad, indistinct). In order to test the validity of the simulated, optimized model, jackknife parameter error estimates were determined and are presented in table 2. 

\begin{center}
\begin{table}
\caption{Results for simulation of hydrous albite 3QMAS spectrum}

\begin{tabular}{lllllll}
\hline
Site \# & $\delta_{iso}^{cs} (Hz/ppm)$ &  & $C_q (MHz)$ & & $\eta_q$ & Rel. Popu
lation \\
&$\mu$ & $\sigma$ & $\mu $ &$\sigma$ &  &  \\
\hline
1 & 8035/61.7 & 172/1.3 & 3.8 & 1.5 & 0.21 & 0.34 \\
2 &  8623/66.2 & 167/1.3 & 2.9 & 0.8 & 0.51 & 0.66 \\
\end{tabular}

\end{table}
\end{center}

\begin{center}
\begin{table}
\caption{Jackknife parameter error estimates for simulation of hydrous albite 3QMAS spectrum}
\begin{tabular}{lllllll}
\hline
Site \# & $\delta_{iso}^{cs}$ (\%) &  & $C_q $(\%) & & $\eta_q$ (\%) & Rel. Popu
lation (\%) \\
& $\mu$ & $\sigma$ & $\mu $ & $\sigma$ &  &  \\ \hline
1 & 1.1& 4.8 & 8.1& 7.8 & 8.6 & 18.0 \\
2 & 0.2& 1.9 & 1.4& 1.7 & 0.2 & 4.0 \\
\end{tabular}
\end{table}
\end{center}

The chemical site with narrow distributions (assigned here to crystalline albite, in good agreement with prior investigations by other workers~\cite{yan89}) has corresponding parameters with least error. This may be attributed to a number of factors, in this case most likely to the lower signal to noise ratio of the disordered region, assigned here to amorphous albite glass.  For chemical sites with larger quadrupole coupling constants, there is also the possibility that due to experimental excitation deficiency, the second order perturbation frequency expression breaks down. Finally, the assumptions of a Gaussian statistical model may be inappropriate for the system in question.  As mentioned earlier, model distributions reflect the underlying stochastic nature of bonding in a disordered material. There is significant evidence~\cite{esp08,cae98,neu04} to suggest that a more general electric field gradient model for disordered systems as probed by $^{27}$Al NMR is given by the Czjzek model~\cite{czj81}. Future work will be devoted to the incorporation of models such as these into the approach outlined here.

\section{Summary}
Theory has been outlined and an application implemented in the C programming language that permits the simulation of an MQMAS spectrum, as a function of underlying parameter distributions. This simulation relies on the use of quasi-Monte Carlo variates to promote convergence and utilizes the OpenMP library to permit execution on SMP machines. Owing to the manner in which random variates are created in the application, the program is amenable to High Throughput Computing (HTC) platforms such as Condor or PBS. In addition, an OCTAVE script implementing a simulated annealing algorithm is used to optimize the simulation, providing reliable estimates of NMR parameters. Finally, theory was outlined and implemented for providing parameter variance estimates using a jackknife approach. In conjunction with the MQMAS experiment, the application described herein enables the characterization of materials which may vary greatly in the degree of underlying chemical and structural order.

\section*{Acknowledgements}
Jeff Nucciarone and the Research Computing and Cyberinfrastructure group at Penn State are acknowledged for their generous assistance and use of computational resources. Marek Pruski provided the MQMAS spinsight pulse sequence used for experiments. Dominique Massiot and Zhehong Gan kindly provided many helpful remarks regarding the preparation of this manuscript. This work has been funded via National Science Foundation grant number CHE 0535656. KTM and MCD acknowledge further funding through the Penn State Center for Environmental Kinetics Analysis, supported by the National Science Foundation through grant CHE 0431328. 

\section*{Appendix A}
Define:
\[clb_0 = -I(I+1)+3/4\]
\[clb_1 = -18I(I+1)+34/4+5\]
\[clb_2 = (r-c)(I(I+1)-3(r^2+rc+c^2))\]
\beq clb_3 = (r-c)(18I(I+1)-34(r^2+rc+c^2)-5)\eeq
then 

\beq\frac{\partial {\cal I}}{\partial \eta_q}=\frac{\partial {\cal I}}{\partial f1_m}\frac{d f1_m}{d \eta_q} + \frac{\partial {\cal I}}{\partial f2_m}\frac{d f2_m}{d \eta_q} \eeq

where:

\[ \frac{d f2_m}{d \eta_q}=\]

\[\frac{clb_1 y^2}{15120 f_0 I^2 (2I-1)^2 } \left \{ \cos ^2\beta \left(140 \cos (4.0 \alpha) \eta_q+60.0 \eta_q -480 \cos (2\alpha)\right)\right.\]

 \[  + \cos ^4\beta\,\left(-70.0\,\cos  \left(4\,\alpha\right)\,\eta_q
-70.0\,\eta_q+420\,\cos \left(2\,\alpha\right) \right) -70.0 \cos(4.0 \alpha) \eta_q\]

\beq \left. -6.0 \eta_q 60.0 \cos(2\alpha)\right \} -\frac{clb_0 y^2 \eta_q}{5 f_0 I^2 (2I-1)^2} \eeq

\[ \frac{d f1_m}{d \eta_q}=\]
\[\frac{-k \cdot clb_1 y^2}{15120 f_0 I^2 (2I-1)^2 } \left \{ \cos ^2\beta \left(140 \cos (4.0 \alpha) \eta_q+60.0 \eta_q -480 \cos (2\alpha)\right)\right.\]

 \[  + \cos ^4\beta\,\left(-70.0\,\cos  \left(4\,\alpha\right)\,\eta_q
-70.0\,\eta_q+420\,\cos \left(2\,\alpha\right) \right) -70.0 \cos(4.0 \alpha) \eta_q\]

\[ \left. -6.0 \eta_q+60.0 \cos(2\alpha)\right \} + \frac{k \cdot clb_0 y^2 \eta_q}{5 f_0 I^2 (2I-1)^2}\]

\[ -\frac{clb_2 y^2}{15120 f_0I^2(2I-1)^2}\left\{ \cos ^2\beta \left(140\,\cos \left(4.0\alpha\right)\eta_q+60.0 \eta_q-480\,\cos (2\alpha)\right) \right.\]

\[  + \cos  ^4\beta\,\left(-70.0\,\cos \left(4\,\alpha\right)\,\eta_q-70.0\,\eta_q+420\, \cos \left(2\,\alpha\right)\right)-70.0\,\cos \left(4.0\,\alpha\right)\,\eta_q\]

\beq \left. -6.0 \eta_q+60.0\,\cos(2\alpha)\right \} +\frac{clb_3 y^2 \eta_q}{5 f_0 I^2(2I-1)^2} \eeq

where $k$ is the shear factor of the MQMAS or STMAS experiment,

\[
\frac{\partial {\cal I}}{\partial f2_m}=
\]

\[{\cal A}_iP_i(x,y)\left(\frac{(f2-f2_m)e^{-\frac{(f2 -f2_m)^2}{2\lambda_2^2}-\frac{( f1-f1_m)^2}{2\lambda_1^2}}\epsilon}{2\pi\lambda_1\lambda_2^3}\right. \]

\beq \left. +\frac{2( f2-f2_m)\lambda_1\lambda_2(1-\epsilon)}{\left(\lambda_1^2+\left(f1- f1_m\right)^2\right) \left(\lambda_2^2+\left( f2-f2_m\right)^2\right)^2}\right)\eeq

\[
\frac{\partial {\cal I}}{\partial f1_m}=
\]
\[{\cal A}_iP_i(x,y)\left(\frac{(f1-f1_m)e^{-\frac{(f2 -f2_m)^2}{2\lambda_2^2}-\frac{( f1-f1_m)^2}{2\lambda_1^2}}\epsilon}{2\pi\lambda_2\lambda_1^3}\right. \]
\beq
   \left. +\frac{2( f1-f1_m)\lambda_1\lambda_2(1-\epsilon)}{\left(\lambda_2^2+\left(f2- f2_m\right)^2\right) \left(\lambda_1^2+\left( f1-f1_m\right)^2\right)^2}\right)\eeq

\[
\frac{\partial {\cal I}}{\partial \lambda_2}= 
\]
\[{\cal A}_iP_i(x,y)\left( -\frac{ e^{-\frac{\left( f2-f2_m \right)^2}{2 \lambda_2^2} -\frac{\left( f1-f1_m\right)^2}{2 \lambda_1^2}} \epsilon} {2 \pi \lambda_1 \lambda_2^2} + \frac{ \left(f2-f2_m\right)^2 e^{-\frac{ \left(f2-f2_m\right)^2} {2 \lambda_2^2}- \frac{\left(f1-f1_m\right)^2} {2 \lambda_1^2}} \epsilon} {2 \pi \lambda_1 \lambda_2^4} \right. \]

\[ + \frac{ \lambda_1 \left(1-\epsilon\right)}{\left(\lambda_1^2+ \left(f1-f1_m\right)^2\right) \left(\lambda_2^2 +\left(f2-f2_m\right)^2\right)} \]

\beq \left. -\frac{2 \lambda_1  \lambda_2^2 \left(1-\epsilon\right)}{\left(\lambda_1^2 +\left(f1-f1_m\right)^2\right) \left(\lambda_2^2+\left(f2-f2_m\right)^2\right)^2}\right)\eeq

\[
\frac{\partial {\cal I}}{\partial \lambda_1}= 
\]
\[{\cal A}_iP_i(x,y)\left( -\frac{ e^{-\frac{\left( f2-f2_m \right)^2}{2 \lambda_2^2} -\frac{\left( f1-f1_m\right)^2}{2 \lambda_1^2}} \epsilon} {2 \pi \lambda_2 \lambda_1^2} + \frac{ \left(f2-f2_m\right)^2 e^{-\frac{ \left(f2-f2_m\right)^2} {2 \lambda_2^2}- \frac{\left(f1-f1_m\right)^2} {2 \lambda_1^2}} \epsilon} {2 \pi \lambda_2 \lambda_1^4} \right. \]

\[ + \frac{ \lambda_2 \left(1-\epsilon\right)}{\left(\lambda_1^2+ \left(f1-f1_m\right)^2\right) \left(\lambda_2^2 +\left(f2-f2_m\right)^2\right)} \]

\beq \left. -\frac{2 \lambda_1^2  \lambda_2 \left(1-\epsilon\right)}{\left(\lambda_1^2 +\left(f1-f1_m\right)^2\right) \left(\lambda_2^2+\left(f2-f2_m\right)^2\right)^2}\right) \eeq

\[\frac{\partial {\cal I}}{\partial \rho} = \]

\[{\cal A}_iP_i(x,y)F_i(f1,f2) \times \]

\beq \left ( \frac{\rho}{(1-\rho^2)} - \frac{\rho(x-\mu_x)^2}{(1-\rho^2)^2\sigma_x} + \frac{(x-\mu_x)(y-\mu_y)}{(1-\rho^2)\sigma_x\sigma_y}+\frac{2\rho^2(x-\mu_x)}{(1-\rho^2)\sigma_x\sigma_y} - \frac{\rho(y-\mu_y)^2}{(1-\rho^2)^2\sigma_y} \right ) \eeq

\[
\frac{\partial {\cal I}}{\partial \sigma_y}=
\]
\[ -\frac{{\cal A}_iF_i(f1,f2)P_i(x,y)}{2 \left(1-\rho^2\right)} \left(\frac{2 \rho \left(x- \mu_x\right) \left(y-\mu_y\right)} {\sigma_x \sigma_y^2} -\frac{2 \left(y-\mu_y\right)^2}{ \sigma_y^3} \right)\]

\beq - \frac{{\cal A}_iF_i(f1,f2)P_i(x,y)}{\sigma_y}\eeq

\[
\frac{\partial {\cal I}}{\partial \mu_y}=
\]

\beq -\frac{{\cal A}_iF_i(f1,f2)P_i(x,y)}{2 \left(1-\rho^2\right)}\left({{2\,\rho\,\left(x- \mu_x\right)}\over{\sigma_x\,\sigma_y}}-{{2\, \left(y-\mu_y\right)}\over{\sigma_y^2}}\right) \eeq

\[
\frac{\partial {\cal I}}{\partial \sigma_x}=
\]

\[ -\frac{{\cal A}_iF_i(f1,f2)P_i(x,y)}{2 \left(1-\rho^2\right)} \left(\frac{2 \rho \left(y- \mu_y\right) \left(x-\mu_x\right)} {\sigma_y \sigma_x^2} -\frac{2 \left(x-\mu_x\right)^2}{ \sigma_x^3} \right) \]

\beq - \frac{{\cal A}_iF_i(f1,f2)P_i(x,y)}{\sigma_x}\eeq

\[
\frac{\partial {\cal I}}{\partial \mu_x}=
\]

\beq -\frac{{\cal A}_iF_i(f1,f2)P_i(x,y)}{2 \left(1-\rho^2\right)}\left({{2\,\rho\,\left(y- \mu_y\right)}\over{\sigma_y\,\sigma_x}}-{{2\, \left(x-\mu_x\right)}\over{\sigma_x^2}}\right) \eeq

\beq \frac{\partial {\cal I}}{\partial {\cal A}}= F_i(f1,f2)P_i(x,y) \eeq

\bibliography{cpc_arxiv}

\end{document}